\documentclass[runningheads]{llncs}
\usepackage[T1]{fontenc}
\usepackage[misc]{ifsym}
\usepackage[colorlinks=true,linkcolor=black,citecolor=blue,urlcolor=blue,]{hyperref}
\usepackage{graphicx}
\usepackage{multirow}

\begin{document}
\title{SegNetr: Rethinking the local-global interactions and skip connections in U-shaped networks}
\titlerunning{SegNetr}
% If the paper title is too long for the running head, you can set
% an abbreviated paper title here
% %
\author{Junlong Cheng\inst{1} \and Chengrui Gao\inst{1} \and
Fengjie Wang\inst{1} \and Min Zhu\inst{1} \textsuperscript{(\Letter)}}
\authorrunning{Junlong Cheng et al.}
% % First names are abbreviated in the running head.
% % If there are more than two authors, 'et al.' is used.
% %
\institute{College of Computer Science, Sichuan University, Chengdu 610065, China\\
\email{zhumin@scu.edu.cn}}
\maketitle              % typeset the header of the contribution
\begin{abstract}
Recently, U-shaped networks have dominated the field of medical image segmentation due to their simple and easily tuned structure. However, existing U-shaped segmentation networks: 1) mostly focus on designing complex self-attention modules to compensate for the lack of long-term dependence based on convolution operation, which increases the overall number of parameters and computational complexity of the network; 
2) simply fuse the features of encoder and decoder, ignoring the connection between their spatial locations. In this paper, we rethink the above problem and build a lightweight medical image segmentation network, called SegNetr. Specifically,  we introduce a novel SegNetr block that can perform local-global interactions dynamically at any stage and with only linear complexity. At the same time, we design a general information retention skip connection (IRSC) to preserve the spatial location information of encoder features and achieve accurate fusion with the decoder features. We validate the effectiveness of SegNetr on four mainstream medical image segmentation datasets, with 59\% and 76\% fewer parameters and GFLOPs than vanilla U-Net, while achieving segmentation performance comparable to state-of-the-art methods. Notably, the components proposed in this paper can be applied to other U-shaped networks to improve their segmentation performance.

\end{abstract}
\section{Introduction}
Medical image segmentation has been one of the key aspects in developing automated assisted diagnosis systems, which aims to separate objects or structures in medical images for independent analysis and processing. 
Normally, segmentation needs to be performed manually by professional physicians, which is time-consuming and error-prone. 
In contrast, developing computer-aided segmentation algorithms can be faster and more accurate for batch processing. 
% 
% about related work
The approach represented by U-Net~\cite{01} is a general architecture for medical image segmentation, which generates a hierarchical feature representation of the image through a top-down encoder path and uses a bottom-up decoder path to map the learned feature representation to the original resolution to achieve pixel-by-pixel classification. 
After U-Net, U-shaped methods based on \textbf{Convolutional Neural Networks} (CNN) have been extended for various medical image segmentation tasks~\cite{02,03,04,05,06,07,08,09}. They either enhance the feature representation capabilities of the encoder-decoder or carefully design the attention module to focus on specific content in the image. 
Although these extensions can improve the benchmark approach, the local nature of the convolution limits them to capturing long-term dependencies, which is critical for medical image segmentation. Recently, segmentation methods based on U-shaped networks have undergone significant changes driven by \textbf{Transformer}~\cite{10,11}.
Chen et al~\cite{12} proposed the first Transformer-based U-shaped segmentation network. 
Cao et al~\cite{13} extended the Swin Transformer~\cite{14} directly to the U-shaped structure. The above methods suffer from high computational and memory cost explosion when the feature map size becomes large.  In addition, some researchers have tried to build \textbf{Hybrid Networks} by combining the advantages of CNN and Transformer, such as UNeXt~\cite{15}, TransFuse~\cite{16}, MedT~\cite{17}, and FAT-Net~\cite{18}. 
Similar to these works, we redesign the window-based local-global interaction and insert it into a pure convolutional framework to compensate for the deficiency of convolution in capturing global features and to reduce the high computational cost arising from self-attention operations.

\textbf{Skip connection} is the most basic operation for fusing shallow and deep features in U-shaped networks. 
Considering that this simple fusion does not fully exploit the information, researchers have proposed some novel ways of skip connection~\cite{19,20,21,22}. 
UNet++~\cite{19} design a series of dense skip connections to reduce the semantic gap between the encoder and decoder sub-network feature maps. 
SegNet~\cite{20} used the maximum pooling index to determine the location information to avoid the ambiguity problem during up-sampling using deconvolution. 
BiO-Net~\cite{21} proposed bi-directional skip connections to reuse building blocks in a cyclic manner. 
UCTransNet~\cite{22} designed a Transformer-based channel feature fusion method to bridge the semantic gap between shallow and deep features.  Our approach focuses on the connection between the spatial locations of the encoder and decoder, preserving more of the original features to help recover the resolution of the feature map in the upsampling phase, and thus obtaining a more accurate segmentation map.

By reviewing the above multiple successful cases based on U-shaped structure, we believe that the efficiency and performance of U-shaped networks can be improved by improving the following two aspects: 
\textbf{(i) local-global interactions.} Often networks need to deal with objects of different sizes in medical images, and local-global interactions can help the network understand the content of the images more accurately. 
\textbf{(ii) Spatial connection between encoder-decoder.} Semantically stronger and positionally more accurate features can be obtained using the spatial information between encoder-decoders. 
Based on the above analysis, this paper rethinks the design of the U-shaped network. 
Specifically, we construct lightweight SegNetr (\textbf{Seg}mentation \textbf{Net}work with T\textbf{r}ansformer) blocks to dynamically learn local-global information over non-overlapping windows and maintain linear complexity. 
We propose information retention skip connection (IRSC), which focuses on the connection between encoder and decoder spatial locations, retaining more original features to help recover the resolution of the feature map in the up-sampling phase. 
In summary, the contributions of this paper can be summarized as follows: 
1) We propose a lightweight U-shape SegNetr segmentation network with less computational cost and better segmentation performance. 
2) We investigate the potential deficiency of the traditional U-shaped framework for skip connection and improve a skip connection with information retention. 
3) When we apply the components proposed in this paper to other U-shaped methods, the segmentation performance obtains a consistent improvement. 

\begin{figure}\centering
\includegraphics[width=\textwidth]{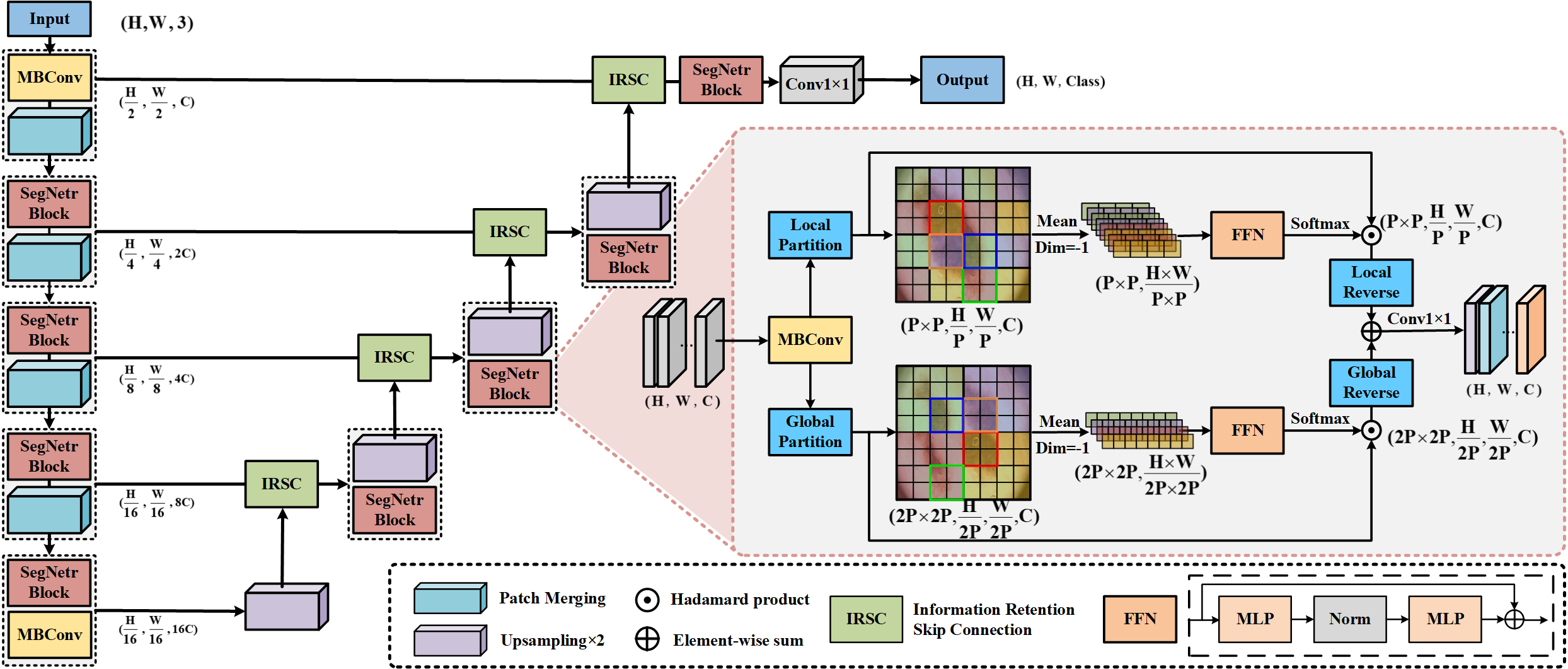}
\caption{Overview of the SegNetr approach. SegNetr blocks interact through parallel local and global branches. IRSC preserves the positional information of encoder features and achieves accurate fusion with decoder features.} \label{fig1}
\end{figure}

\section{Method}
As shown in Fig.~\ref{fig1}, SegNetr is a hierarchical U-shaped network with important components including SegNetr blocks and IRSC. 
To make the network more lightweight, we use MBConv~\cite{24} as the base convolutional building block. 
SegNetr blocks implement dynamic local-global interaction in the encoder and decoder stages. 
Patch merging~\cite{14} is used to reduce the resolution by a factor of two without losing the original image information. 
IRSC is used to fuse encoder and decoder features, reducing the detailed information lost by the network as the depth deepens. 
Note that by changing the number of channels, we can get the smaller version of SegNetr-S (C=32) and the standard version of SegNetr (C=64). 
Next, we will explain in detail the important components in SegNetr.

\subsection{SegNetr Block}
The self-attention mechanism with global interactions is one of the keys to Transformer's success, but computing the attention matrix over the entire space requires a quadratic complexity. Inspired by the window attention method~\cite{14,23}, we construct SegNetr blocks that require only linear complexity to implement local-global interactions. Let the input feature map be $X\in R^{H\times W\times C}$. We first extract the feature $X_{MBConv}\in R^{H\times W\times C}$ using MBConv~\cite{24}, which provides non-explicit position encoding compared to the usual convolutional layer.

\textbf{Local interaction} can be achieved by calculating the attention matrix of non-overlapping small patches ($P$ for patch size). First, we divide $X_{MBConv}$ into a series of patches $(\frac{H\times W}{P\times P},P,P,C)$ that are spatially continuous (Fig.~\ref{fig1} shows the patch size for $P = 2$) using a computationally costless local partition (LP) operation. Then, we average the information of the channel dimensions and flatten the spatial dimensions to obtain $(\frac{H\times W}{P\times P},P\times P)$, which is fed into the FFN~\cite{11} for linear computation. Since the importance of the channel aspect is weighed in MBConv~\cite{24}, we focus on the computation of spatial attention when performing local interactions. Finally, we use Softamx to obtain the spatial dimensional probability distribution and weight the input features $X_{MBConv}$. This approach is not only beneficial for parallel computation, but also focuses more purely on the importance of the local space.

Considering that local interactions are not sufficient and may have under-fitting problems, we also design parallel \textbf{global interaction} branches. First, we use the global partition (GP) operation to aggregate non-contiguous patches on the space. GP adds the operation of window displacement to LP with the aim of changing the overall distribution of features in space (The global branch in Fig.~\ref{fig1} shows the change in patch space location after displacement). The displacement rules are one window to the left for odd patches in the horizontal direction (and vice versa for even patches to the right), and one window up for odd patches in the vertical direction (and vice versa for even patches down). Note that the displacement of patches does not have any computational cost and only memory changes occur. Compared to the sliding window operation of~\cite{14}, our approach is more global in nature. Then, we decompose the spatially shifted feature map into $2P$ $(\frac{H\times W}{2P\times 2P},2P,2P,C)$ patches and perform global attention computation (similar to the local interaction branch). Even though the global interaction computes the attention matrix over a larger window relative to the local interaction operation, the amount of computation required is much smaller than that of the standard self-attention model.

\begin{figure}\centering
\includegraphics[width=0.9\textwidth]{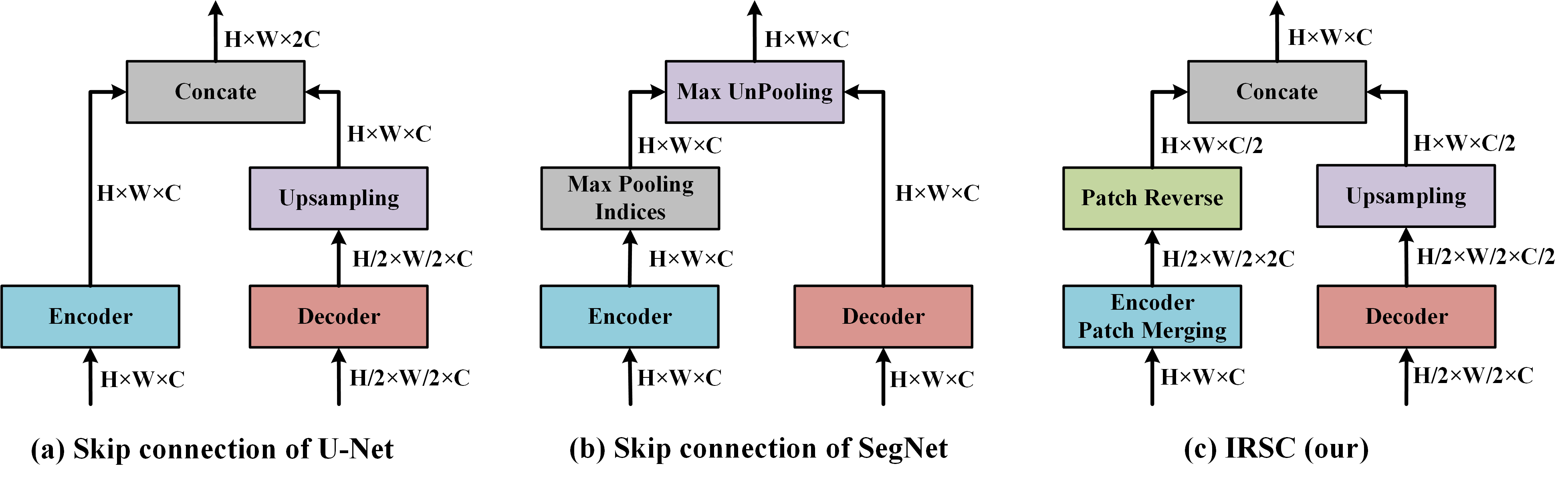}
\caption{Comparison of skip connections of U-Net, SegNet and our SegNetr. Our method does not incorporate redundant computable modules, but the patch reverse (PR) provides spatial location information.} \label{fig2}
\end{figure}

The local and global branches are finally fused by weighted summation, before which the feature map shape needs to be recovered by LP and GP reversal operations (i.e., local reverse (LR) and global reverse (GR)). In addition, our approach also employs efficient designs of Transformer, such as Norm, feed-forward networks (FFN) and residual connections. Most Transformer models use fixed-size patches~\cite{11,12,13,14,24}, but this approach limits them to focus on a wider range of regions in the early stages. This paper alleviates this problem by applying dynamically sized patches. In the encoder stage, we compute local attention using patches of $(8, 4, 2, 1)$ in turn, and the global branch expands patches to the size of $(16, 8, 4, 2)$. To reduce the hyper-parameter setting, the patches of the decoder stage are of the same size as the encoder patches of the corresponding stage. %This setting is consistent with human visual perception of images, i.e., attention is focused on larger areas first when the resolution is larger and shifts to smaller and more interesting areas as the resolution gradually decreases.

\subsection{Information Retention Skip Connection}
Fig.~\ref{fig2} shows three different types of skip connections. U-Net splices the channel dimensions at the corresponding stages of the encoder and decoder, allowing the decoder to retain more high-resolution detail information when performing up-sampling. SegNet assists the decoder to recover the feature map resolution by retaining the position information of the down-sampling process in the encoder. We design the IRSC to consider both of these features, i.e., to preserve the location information of encoder features while achieving the fusion of shallow and deep features. Specifically, the patch merging (PM) operation in the encoder reduces the resolution of the input feature map $X_{in}\in R^{H\times W\times C}$ to twice the original one, while the channel dimension is expanded to four times the original one to obtain $X_{PM}\in R^{\frac{H}{2}\times \frac{W}{2}\times 4C}$. The essence of the PM operation is to convert the information in the spatial dimension into a channel representation without any computational cost and retaining all the information of the input features. The patch reverse (PR) in IRSC is used to recover the spatial resolution of the encoder, and it is a reciprocal operation with PM. We alternately select half the number of channels of $X_{PM}$ (i.e., ${\frac{H}{2}\times \frac{W}{2}\times 2C}$) as the input of PR, which can reduce the redundant features in the encoder on the one hand and align the number of feature channels in the decoder on the other hand. PR reduces the problem of information loss to a large extent compared to traditional up-sampling methods, while providing accurate location information. Finally, the output features $X_{PR}\in R^{H\times {W}\times \frac{C}{2}}$ of PR are fused with the up-sampled features of the decoder for the next stage of learning.

\section{Experiments and Discussion}
\noindent\textbf{Datasets.} To verify the validity of SegNetr, we selected four datasets, ISIC2017~\cite{25}, PH2~\cite{26}, TNSCUI~\cite{27} and ACDC~\cite{28}, for benchmarking. ISIC2017 consists of 2000 training images, 200 validation images, and 600 test images. The PH2 and ISIC2017 tasks are the same, but this dataset contains only 200 images without any specific test set, so we use a five-fold cross-validation approach to validate the different models. The TNSCUI dataset has 3644 ultrasound images of thyroid nodules, which we randomly divided into a 6:2:2 ratio for training, validation, and testing. The ACDC contains Cardiac MRI images from 150 patients, and we obtained a total of 1489 slice images from 150 3D images, of which 951 were used for training and 538 for testing. Unlike the three datasets mentioned above, the ACDC dataset contains three categories: left ventricle (LV), right ventricle (RV), and myocardium (Myo). We use this dataset to explore the performance of different models for multi-category segmentation.\par

\noindent\textbf{Implementation details.} We implement the SegNetr method based on the PyTorch framework by training on an NVIDIA 3090 GPU with 24 GB of memory. Use the Adam optimizer with a fixed learning rate of 1e-4. All networks use a cross-entropy loss function and an input image resolution of 224 × 224, and training is stopped when 200 epochs are iteratively optimized. We use the source code provided by the authors to conduct experiments with the same dataset, and data enhancement strategy. In addition, we use the IoU and Dice metrics to evaluate the segmentation performance, while giving the number of parameters and GFLOPs for the comparison models.\par

\begin{table}\centering
\caption{Quantitative results on ISIC2017 and PH2 datasets.}\label{tab1}
\setlength{\tabcolsep}{1.5mm}{
\begin{tabular}{l|cc|cc|cc}
\hline
\multirow{2}{*}{Network} & \multicolumn{2}{c|}{ISIC2017} & \multicolumn{2}{c|}{PH2}  & \multirow{2}{*}{Params} & \multirow{2}{*}{GFLOPs} \\ \cline{2-5}
                         & IoU           & Dice          & IoU         & Dice        &                         &                         \\ \hline
U-Net~\cite{01}             & 0.736         & 0.825         & 0.878±0.025 & 0.919±0.045 & 29.59 M                 & 41.83                   \\
SegNet~\cite{20}            & 0.696         & 0.821         & 0.880±0.020 & 0.934±0.012 & 17.94 M                 & 22.35                   \\
UNet++~\cite{19}           & 0.753         & 0.840         & 0.883±0.013 & 0.936±0.008 & 25.66 M                 & 28.77                   \\
FAT-Net~\cite{18}            & 0.765         & 0.850         & \textcolor{green}{0.895±0.019} & \textcolor{green}{0.943±0.011} & 28.23 M                 & 42.83                   \\
ResGANet~\cite{05}          & 0.764         & \textcolor{red}{0.862}         & —           & —           & 39.21 M                 & 65.10                   \\
nnU-Net~\cite{29}            & 0.760         & 0.843         & —           & —           & —                       & —                       \\
Swin-UNet~\cite{13}         & \textcolor{green}{0.767}         & 0.850         & 0.872±0.022 & 0.927±0.014 & 25.86 M                 & 5.86                    \\
TransUNet~\cite{12}        & \textcolor{red}{0.775}         & 0.847         & 0.887±0.020 & 0.937±0.012 & 88.87 M                 & 24.63                   \\
UNeXt-L~\cite{15}           & 0.754         & 0.840         & 0.884±0.021 & 0.936±0.013 & \textcolor{green}{3.80 M}                  & \textcolor{red}{1.08}                    \\ \hline
SegNetr-S                & 0.752         & 0.838         & 0.889±0.018 & 0.939±0.011 & \textcolor{red}{3.60 M}                  & \textcolor{green}{2.71}                    \\
SegNetr                  & \textcolor{red}{0.775}         & \textcolor{green}{0.856}         & \textcolor{red}{0.905±0.023} & \textcolor{red}{0.948±0.014} & 12.26 M                 & 10.18                   \\ \hline
\end{tabular}}
\end{table}

\subsection{Comparison with State-of-the-arts}

\noindent\textbf{ISIC2017 and PH2 Results.} As shown in Table.~\ref{tab1}, we compared SegNetr with the baseline U-Net and eight other state-of-the-art methods~\cite{05,12,13,15,18,19,20,29}. On the ISIC2017 dataset, SegNetr and TransUNet obtained the highest IoU (0.775), which is 3.9\% higher than the baseline U-Net. Even SegNetr-S with a smaller number of parameters can obtain a segmentation performance similar to that of its UNeXt-L counterpart. By observing the experimental results of PH2, we found that the Transformer-based method Swin-UNet segmentation has the worst performance, which is directly related to the data volume of the target dataset. Our method obtains the best segmentation performance on this dataset and keeps the overhead low. Although we use an attention method based on window displacement, the convolutional neural network has a better inductive bias, so the dependence on the amount of data is smaller compared to Transformer-based methods such as Swin-UNet or TransUNet.\par

% Please add the following required packages to your document preamble:
% \usepackage{multirow}
\begin{table}\centering
\caption{Quantitative results on TNSCUI and ACDC datasets.}\label{tab2}
\setlength{\tabcolsep}{1.1mm}{
\begin{tabular}{l|c|cccc|cc}
\hline
\multirow{2}{*}{Network} & TNSCUI       & \multicolumn{3}{c}{ACDC / IoU}             & \multicolumn{1}{c|}{\multirow{2}{*}{\begin{tabular}[c]{@{}c@{}}Average\\ IoU (Dice)\end{tabular}}} & \multirow{2}{*}{Params} & \multirow{2}{*}{GFLOPs} \\ \cline{2-5}
                         & IoU (Dice)   & RV           & Myo          & LV           &                                                                               &                         &                         \\ \hline
U-Net~\cite{01}             & 0.718 (0.806) & 0.743 & 0.717 & 0.861 & 0.774 (0.834)                                                                  & 29.59 M                 & 41.83                   \\
SegNet~\cite{20}            & 0.726 (0.819) & 0.738 & 0.720 & 0.864 & 0.774 (0.836)                                                                  & 17.94 M                 & 22.35                   \\
FAT-Net~\cite{18}            & \textcolor{green}{0.751} (\textcolor{green}{0.842}) & 0.743 & 0.702 & 0.859 & 0.768 (0.834)                                                                  & 28.23 M                 & 42.83                   \\
Swin-UNet~\cite{13}         & 0.744 (0.835) & \textcolor{green}{0.754} & 0.722 & 0.865 & \textcolor{green}{0.780} (\textcolor{green}{0.843})                                                                  & 25.86 M                 & 5.86                    \\
TransUNet~\cite{12}        & 0.746 (0.837) & 0.750 & 0.715 & \textcolor{green}{0.866} & 0.777 (0.838)                                                                  & 88.87 M                 & 24.63                   \\
EANet~\cite{30}             & \textcolor{green}{0.751} (0.839) & 0.742 & \textcolor{green}{0.732} & 0.864 & 0.779 (0.839)                                                                  & 47.07 M                 & 98.63                   \\
UNeXt~\cite{15}            & 0.655 (0.749) & 0.697 & 0.646 & 0.814 & 0.719 (0.796)                                                                  & \textcolor{red}{1.40 M}                  & \textcolor{red}{0.44}                    \\
UNeXt-L~\cite{15}           & 0.693 (0.794) & 0.719 & 0.675 & 0.840 & 0.744 (0.815)                                                                  & 3.80 M                  & \textcolor{green}{1.08}                    \\ \hline
SegNetr-S                & 0.707 (0.804) & 0.723 & 0.692 & 0.845 & 0.753 (0.821)                                                                  & \textcolor{green}{3.60 M}                  & 2.71                    \\
SegNetr                  & \textcolor{red}{0.767} (\textcolor{red}{0.850}) & \textcolor{red}{0.761} & \textcolor{red}{0.738} & \textcolor{red}{0.872} & \textcolor{red}{0.791} (\textcolor{red}{0.847})                                                                  & 12.26 M                 & 10.18                   \\ \hline
\end{tabular}}
\end{table}

\noindent\textbf{TNSCUI and ACDC Results.} As shown in Table~\ref{tab2}, SegNetr's IoU and Dice are 1.6\% and 0.8 higher than those of the dual encoder FATNet, respectively, while the GFLOPs are 32.65 less. In the ACDC dataset, the left ventricle is easier to segment, with an IoU of 0.861 for U-Net, but 1.1\% worse than SegNetr. The myocardium is in the middle of the left and right ventricles in an annular pattern, and our method is 0.6\% higher IoU than the EANet that focuses on the boundary segmentation mass. In addition, we observe the segmentation performance of the four networks UNeXt, UNeXt-L, SegNetr-S and SegNetr to find that the smaller parameters may limit the learning ability of the network. The proposed method in this paper shows competitive segmentation performance on all four datasets, indicating that our method has good generalization performance and robustness. Additional qualitative results are in the supplementary.\par

\begin{figure}\centering
\includegraphics[width=1.0\textwidth]{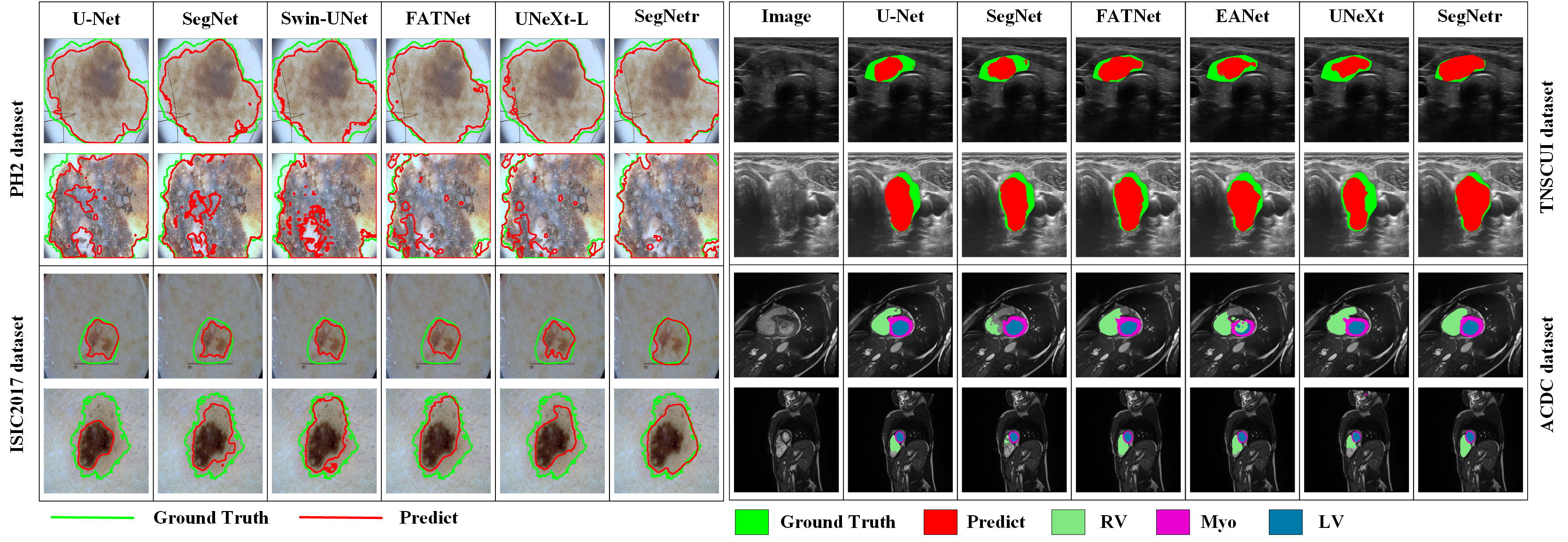}
\caption{Qualitative experimental results of different methods on four datasets.} \label{fig3}
\end{figure}

In addition, Fig. 3 provides qualitative examples that demonstrate the effectiveness and robustness of our proposed method. The results show that SegNetr is capable of accurately describing skin lesions with less data, and achieves multi-class segmentation with minimized under-segmentation and over-segmentation.

\subsection{Ablation Study}
\noindent\textbf{Effect of local-global interactions.} The role of local-global interactions in SegNetr can be understood from Table.~\ref{tab3}. The overall parameters of the network are less when there is no local or global interaction, but the segmentation performance is also greatly affected. With the addition of local or global interactions, the segmentation performance of the network for different categories is improved. In addition, similar performance can be obtained by running the local-global interaction modules in series and parallel, but the series connection leads to lower computational efficiency and affects the running speed.\par

\begin{table}\centering
\caption{Ablation study of local-global interactions on the ACDC dataset.}\label{tab3}
\setlength{\tabcolsep}{0.8mm}{
\begin{tabular}{l|cccc|cc}
\hline
\multirow{2}{*}{Settings} & \multicolumn{3}{c}{ACDC}                   & \multirow{2}{*}{\begin{tabular}[c]{@{}c@{}}Average\\ IoU (Dice)\end{tabular}} & \multirow{2}{*}{Params} & \multirow{2}{*}{GFLOPs} \\ \cline{2-4}
                             & RV           & Myo          & LV           &                                                                                     &                         &                         \\ \hline
Without                      & 0.750(0.799) & 0.720(0.816) & 0.861(0.897) & 0.777(0.837)                                                                        & 10.93 M                 & 9.75                    \\
Only local                   & 0.753(0.800) & 0.733(0.825) & 0.868(0.904) & 0.785(0.843)                                                                        & 12.22 M                 & 10.18                   \\
Only global                  & 0.756(0.803) & 0.734(0.827) & \textcolor{red}{0.875}(\textcolor{red}{0.909}) & 0.788(0.846)                                                                        & \textcolor{red}{11.64 M}                 & \textcolor{red}{10.17}                   \\
Series                       & \textcolor{red}{0.761}(\textcolor{red}{0.809}) & 0.732(0.824) & 0.871(0.907) & 0.788(0.846)                                                                        & 12.26 M                 & 10.18                   \\
Parallel                     & \textcolor{red}{0.761}(0.807) & \textcolor{red}{0.738}(\textcolor{red}{0.828}) & 0.872(0.907) & \textcolor{red}{0.791}(\textcolor{red}{0.847})                                                                        & 12.26 M                 & 10.18                   \\ \hline
\end{tabular}}
\end{table}

\begin{table}\centering
\caption{Ablation study of patch size (left) and IRSC (right) on TNSCUI and ISIC2017 datasets.}\label{tab4}
\setlength{\tabcolsep}{0.9mm}{
\begin{tabular}{l|cc|lc|l|l|cc}
\cline{1-5} \cline{7-9}
\multirow{2}{*}{Patch size} & \multicolumn{2}{c|}{TNSCUI} & \multirow{2}{*}{Params} & \multirow{2}{*}{GFLOPs} &  & \multirow{2}{*}{\begin{tabular}[c]{@{}l@{}}Network\\ +IRSC\end{tabular}} & \multicolumn{2}{c}{ISIC2017}  \\ \cline{2-3} \cline{8-9} 
                            & IoU          & Dice         &                         &                         &  &                                                                          & IoU           & Dice          \\ \cline{1-5} \cline{7-9} 
(2,2,2,2)                   & 0.751        & 0.835        & 54.34 M                 & 10.38                   &  & UNeXt-L                                                                  & \textcolor{red}{0.760}(+0.6\%) & 0.843(+0.3\%) \\
(4,4,4,2)                   & 0.762        & 0.841        & 14.32 M                 & 10.22                   &  & U-Net                                                                    & 0.744(+0.8\%) & 0.839(\textcolor{red}{+1.4\%}) \\
(8,4,4,2)                   & 0.762        & 0.843        & \textcolor{red}{11.96 M}                 & \textcolor{red}{10.18}                   &  & UNet++                                                                  & 0.763(+1.0\%) & \textcolor{red}{0.845}(+0.5\%) \\
(8,4,2,1)                   & \textcolor{red}{0.767}        & \textcolor{red}{0.850}        & 12.26 M                 & \textcolor{red}{10.18}                   &  & SegNet                                                                   & 0.712(\textcolor{red}{+1.6\%}) & 0.829(+0.8\%) \\ \cline{1-5} \cline{7-9} 
\end{tabular}}
\end{table}

\noindent\textbf{Effect of patch size.} As shown in Table.~\ref{tab4} (left), different patch size significantly affects the efficiency and parameters of the model. The number of parameters reaches 54.34 M when patches of size 2 are used in each phase, which is an increase of 42.08 M compared to using dynamic patches of size (8, 4, 2, 1).  Based on this ablation study, we recommend the use of $[\frac{Resolution}{14}]$ patches size at different stages.\par

\noindent\textbf{Effect of IRSC.} Table.~\ref{tab4} (right) shows the experimental results of replacing the skip connections of UNeXt, U-Net, U-Net++, and SegNet with IRSC. These methods get consistent improvement with the help of IRSC, which clearly shows that IRSC is useful.\par

\section{Conclusion}
In this study, we introduce a novel framework SegNetr for medical image segmentation, which achieves segmentation performance improvement by optimizing local-global interactions and skip connections. 
Specifically, the SegNetr block implements dynamic interactions based on non-overlapping windows using parallel local and global branches, and IRSC enables more accurate fusion of shallow and deep features by providing spacial information. 
We evaluated the proposed method using four medical image datasets, and extensive experiments showed that SegNetr is able to obtain challenging experimental results while maintaining a small number of parameters and GFLOPs.
The proposed framework is general and flexible that we believe it can be easily extended to other U-shaped networks.

%
% ---- Bibliography ----
%
% BibTeX users should specify bibliography style 'splncs04'.
% References will then be sorted and formatted in the correct style.
%
% \bibliographystyle{splncs04}
% \bibliography{mybibliography}
%

\end{document}